\documentclass[aps,prl,reprint,groupedaddress,nofootinbib]{revtex4-1}
\usepackage{graphicx, color, soul}
\usepackage{siunitx}
\usepackage{amsmath}
\usepackage{braket}
\usepackage{color,soul,url}
\usepackage{float}

\begin{document}
\title{Observation of a Charge-2 Photonic Weyl Point in the Infrared}
\author{Sachin Vaidya$^{1}$, Jiho Noh$^{1}$, Alexander Cerjan$^{1}$, Mikael C. Rechtsman$^{1}$$^*$}
\affiliation{
 $^1$Department of Physics, The Pennsylvania State University, University Park, PA 16802, USA
}

\def\thefootnote{*}\footnotetext{mcrworld@psu.edu}

\date{\today}

\begin{abstract}
Weyl points are robust point degeneracies in the band structure of a periodic material, which act as monopoles of Berry curvature. They have been at the forefront of research in three-dimensional topological materials as they are associated with novel behavior both in the bulk and on the surface. Here, we present the experimental observation of a charge-2 photonic Weyl point in a low-index-contrast photonic crystal fabricated by two-photon polymerization. The reflection spectrum obtained via Fourier-transform infrared (FTIR) spectroscopy closely matches simulations and shows two bands with quadratic dispersion around a point degeneracy.
\end{abstract}
\maketitle

Over the past decade, topological materials have been studied extensively in the hopes of harnessing properties such as protection from defects and scatter-free transport. The simplest topological systems in three dimensions are Weyl materials, which possess a set of point degeneracies in momentum space that act as sources of Berry curvature \cite{berry}, and are thus topologically protected against perturbations to the system. In this sense, Weyl points are analogous to magnetic monopoles in momentum space and are associated with a quantized topological charge. A direct consequence of their non-zero charge is that perturbations that preserve periodicity cannot cause a gap to open at the Weyl point but merely move it around in the band structure. Moreover, these materials exhibit remarkable surface states that lie on Fermi arcs connecting Weyl points of opposite charges. The necessary conditions for the existence of Weyl points are very general and only require breaking either inversion symmetry, time-reversal symmetry or both. Since this can happen in a wide range of systems, Weyl points have been shown to exist in solids \cite{weylhasan, weylsolids1, weylsolids2, weylsolids3, weylsolids4, weylsolids5, weylsolids6, weylsolids7}, microwave \cite{linglu2,linglu} and optical photonic crystals (PhC) \cite{mingu}, optical waveguide arrays \cite{weyljiho, WERalex}, circuit-based systems \cite{weylsimon}, mechanical crystals \cite{weylmechanical}, phononic crystals \cite{weylacoustic,weylacoustic2}, metamaterials \cite{metametarials1}, magnetized plasmas \cite{magplasma1} and can also be realized using synthetic dimensions \cite{weylsynthetic}.

If a Weyl material has additional spatial symmetries, it is possible for multiple Weyl points of the same charge and between the same two bands to accumulate at high symmetry points in the Brillouin zone leading to Weyl points with a higher topological charge \cite{weylbernevig, weylctchan}. For example, a parity-breaking chiral woodpile PhC made of stacked layers of dielectric rods (n = 3.4) has been predicted to have charge-2 Weyl points \cite{takahashi} that form due to the presence of a screw symmetry. This symmetry pins the location of the $\pm 2$ charges to high symmetry points in the Brillouin zone of the PhC \cite{weylctchan}, which implies that as long as the required spatial symmetries are present, these charge-2 Weyl points must continue to exist regardless of index contrast. While other studies report findings of higher-charged Weyl points, their implementations are limited to macroscopic metallic PhCs that work in microwave frequencies \cite{weylcharge2_microwave} and acoustic structures \cite{weylcharge2_phonon}.

In this letter, we report the experimental observation of a charge-2 Weyl point in the mid-infrared regime in a low-index (n = 1.52) chiral woodpile PhC fabricated by two-photon polymerization and characterized using angle-resolved Fourier-transform infrared (FTIR) spectroscopy. For such a low index contrast, there is only an incomplete bandgap surrounding the Weyl point. Nevertheless, we show that this is sufficient to directly observe the dispersion features associated with the Weyl point in the reflection spectrum of the PhC. Furthermore, we numerically show the existence of topological surface resonances, akin to Fermi arc states, associated with Weyl points that exist in the absence of a bandgap. The prospect of lowering the index contrast requirements for topological photonic structures is of importance since it can allow for readily accessible fabrication techniques such as colloidal self-assembly and two-photon polymerization as well as wider material choices.

\begin{figure*}
\centering
\includegraphics[scale = 0.18]{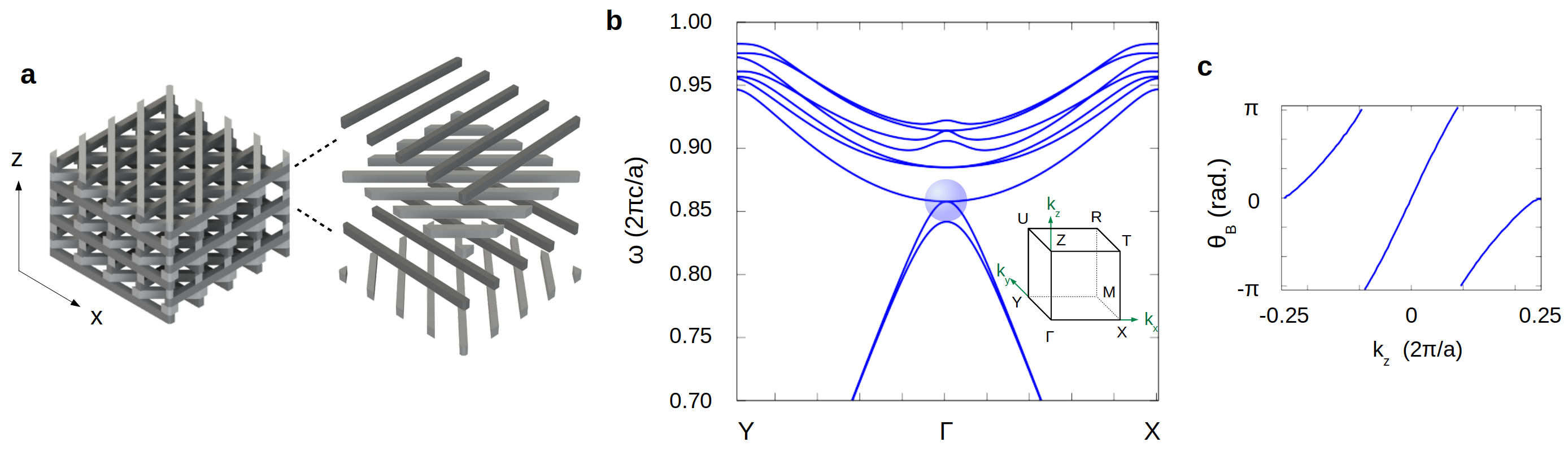}
\caption{\textbf{Chiral woodpile PhC and corresponding band structure supporting a charge-2 Weyl point.} (a) The chiral woodpile structure made by stacking layers of rods with an in-plane rotation of 45$^{\circ}$ between layers. (b) Band structure along $Y$$-\Gamma-$$X$ showing bands 3 to 11. The Weyl point of interest is the degeneracy between bands 4 and 5 at $\Gamma$ (blue circle). The 3D Brillouin zone of the PhC is shown in the inset. (c) Berry phase plot for band 4 around the $\Gamma$ point. The double winding indicates that the degeneracy has a topological charge of 2.}
\label{fig:figure1}
\end{figure*}

The specific PhC we use to realize a charge-2 Weyl point is a chiral woodpile as shown in Fig. \ref{fig:figure1} (a), whose 3D unit cell consists of four layers of rectangular rods stacked on top of each other with a relative 45$^{\circ}$ in-plane rotation. The width and height of the rods is $0.175a$ and $0.25a$ respectively, where $a$ is the lattice constant in all three directions. The spacing between the rods in the 45$^{\circ}$ and 135$^{\circ}$ layers is $a/\sqrt{2}$. The rods have a dielectric constant ($\varepsilon_{\mathrm{rods}}$) of 2.31, which corresponds to the material used in the experiment. Using these parameters, the band structure of our PhC is numerically calculated using MPB \cite{MPB}, and bands 3 to 11 along $Y$$-\Gamma-$$X$ are shown in Fig. \ref{fig:figure1} (b). The degeneracy at $\Gamma$ between bands 4 and 5 is the Weyl point of interest. These bands along $\Gamma-Z$ direction in the Brillouin zone are very close in frequency but convergence tests show that the degeneracy only occurs at the $\Gamma$ point. To confirm the topological nature of this degeneracy, we directly calculate its topological charge (Chern number) by using a discrete algorithm to compute Berry phase ($\theta_B$) given in Ref. \cite{resta}. The phase is calculated using magnetic field eigenmodes from MPB on contours that are defined by constant $k_z$ and enclose the Weyl point. The topological charge of the Weyl point is the number of times the phase winds around as a function of $k_z$ as shown in Fig. \ref{fig:figure1} (c). (see supplementary material for full band structure and details on Berry phase calculation). A second Weyl point of charge $-2$ and between the same two bands is located at the $R$ point in the Brillouin zone, which exhibits double winding of Berry phase in the opposite sense. Since this Weyl point is located below the light line of air and as such is inaccessible without the use of gratings or high-index fluids, we choose to measure the Weyl point at $\Gamma$ in our experiment.
\begin{figure}[t]
\centering
\includegraphics[scale = 0.22]{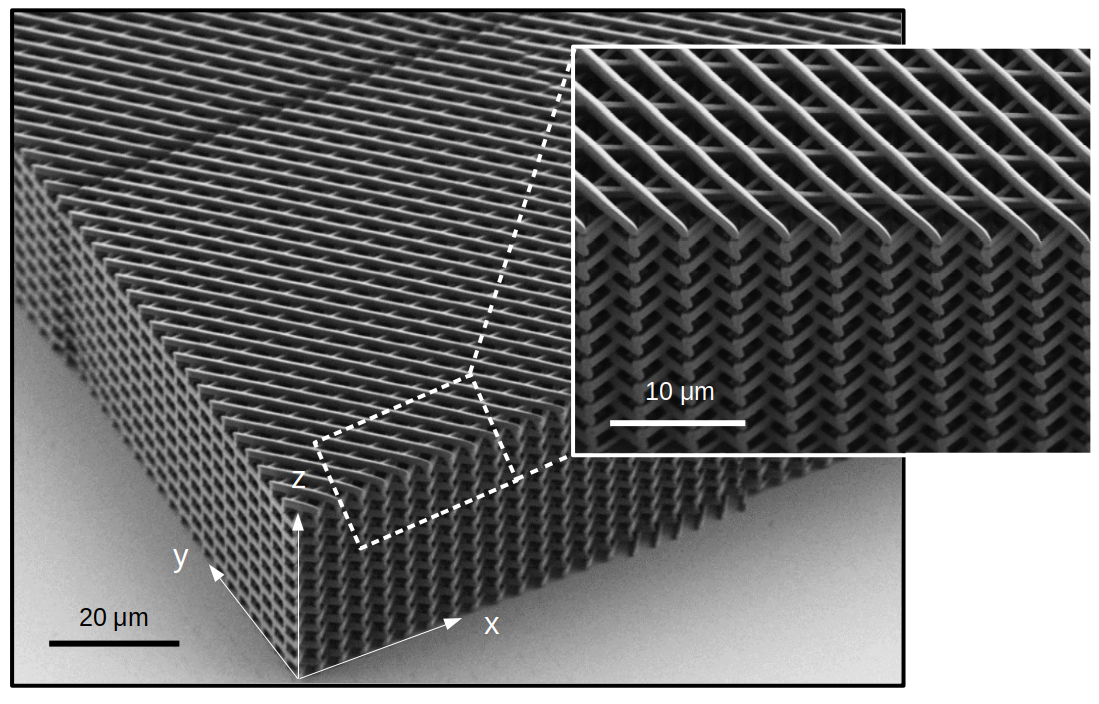}
\caption{\textbf{Fabricated chiral woodpile PhC.} FESEM image of a parity-breaking chiral woodpile fabricated using two-photon polymerization.}
\label{fig:figure2}
\end{figure} 
\begin{figure*}[t]
\centering
\includegraphics[scale = 0.355]{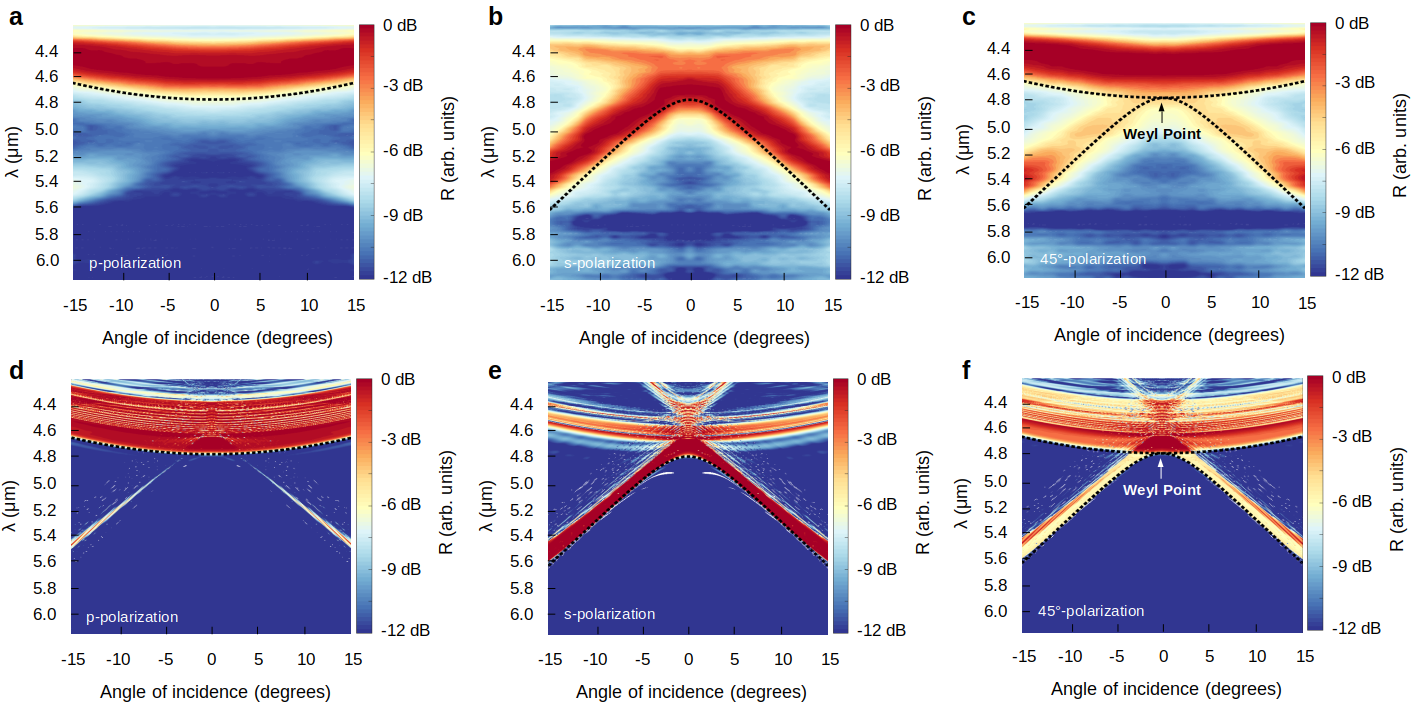}
\caption{\textbf{Observation of the charge-2 Weyl point in the reflection spectra of the chiral woodpile PhC.} (a), (b), (c) Experimentally measured angle-resolved FTIR reflection spectra for 90$^{\circ}$ (p-), 0$^{\circ}$ (s-) and 45$^{\circ}$  polarizations respectively. The dotted lines in all plots are Weyl bulk bands from MPB. (d), (e), (f) S$^4$ (RCWA) simulation of the angle-resolved reflection spectra for all three polarizations.}
\label{fig:figure3}
\end{figure*}

For the experiment, we fabricate the chiral woodpile PhC by 3D lithography (direct laser writing, DLW \cite{exp1}) using a DLW instrument (Photonic Professional GT, Nanoscribe GmbH), which employs two-photon-polymerization of a liquid negative tone photoresist (IP-DIP, Nanoscribe GmbH) (n = 1.52). Here, we use DLW in dip-in configuration \cite{exp2}, where the microscope objective (63x, NA=1.4) for DLW is dipped into the photoresist applied below the fused-silica substrate. The 3D structure is then written layer by layer starting from the bottom surface of the substrate by moving the microscope objective in z-direction. The structure is developed in propylene glycol monomethyl ether acetate (PGMEA) for 20 minutes and then developed in isopropanol for 3 minutes. The thickness of the rods is controlled by varying number of superpositions of individual laser writing for each rod \cite{exp3}. The sample has a lattice constant of 4 $\mu$m and rod width and height equal to 700 nm and 1 $\mu$m respectively. We fabricate 20 unit cells in the z-direction (80 layers) and 250 unit cells in x and y directions. An image of one such sample taken by field emission scanning electron microscopy (FESEM) is shown in Fig. \ref{fig:figure2}.

Reflection measurements are carried out using the Bruker Vertex 80 FTIR spectrometer at wavelengths ranging from $4.2-6$ $\mu$m with the frequency resolution set to 8 cm$^{-1}$. A variable angle reflection accessory (Seagull) is used to observe the angle-resolved spectra from 0$^{\circ}$ to 15$^{\circ}$ along the $\Gamma-$X direction with a resolution of 1$^{\circ}$. The data for $-15^{\circ}$ to $0^{\circ}$ is identical to the data for the corresponding positive angles since the structure is invariant under 180$^{\circ}$ rotational symmetry about the z-axis and hence we only measure data for positive angle and reflect it for negative angles. A polarizer is used at the output port to obtain spectra for 0$^{\circ}$ (s-), 45$^{\circ}$ and 90$^{\circ}$ (p-) polarizations. The spectrum is normalized to the maximum value of reflection for each angle and interpolated. The results from the experiment are shown in  Fig. \ref{fig:figure3} (a), (b) and (c). We also numerically simulate the reflection spectrum using rigorous coupled-wave analysis (RCWA) as implemented in S$^4$ \cite{S4} where periodic boundary conditions are imposed in the lateral directions and the structure is finite in the z direction having 20 unit cells. The simulation results are shown in Fig. \ref{fig:figure3} (d), (e) and (f) for comparison. As can be seen, the simulated and experimentally obtained spectra match very well and show sharp quadratic boundaries separating reflective and transmissive regions that match each of the two relevant bulk Weyl bands in orthogonal polarizations. As expected, the measurement for 45$^{\circ}$ polarization is a superposition of measurements for s- and p-polarizations and shows clear signatures from both bands forming the charge-2 Weyl point at the degeneracy. Since the PhC has low index-contrast, there is no complete band gap around the Weyl point, and therefore this relatively unobscured direct observation warrants further explanation.

\begin{figure*}[t]
\centering
\includegraphics[scale = 0.22]{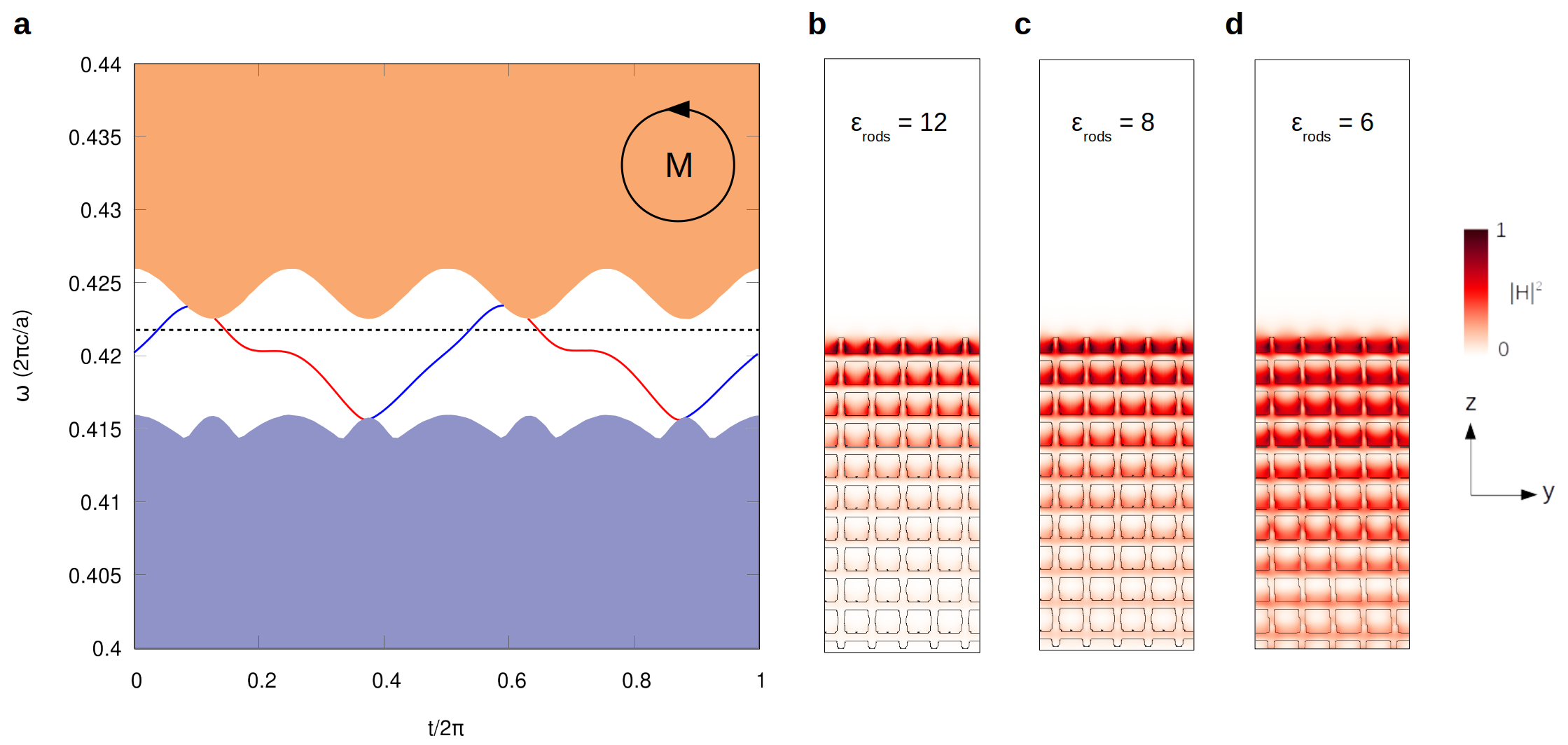}
\caption{\textbf{Surface states associated with the Weyl point at \boldmath$R$} (a) The $k_z$-projected band structure for the truncated chiral woodpile PhC ($\varepsilon_{\mathrm{rods}} = 12$, rod width = 0.175a, rod height = 0.25a) along a loop enclosing the M point with radius $0.15$ ($2\pi/a$) and parametric angle $0\le t/2\pi\le 1$. Solid colors are projections of bulk bands, the solid red and blue lines are surface states localized to the top and bottom surfaces respectively and the black dotted line marks the frequency of the Weyl point. (b)-(c) Magnetic field intensity of the surface state at $t/2\pi = 0.25$ for $\varepsilon_{\mathrm{rods}} = 12$ and $8$ respectively. (d) Magnetic field intensity of the surface resonance for $\varepsilon_{\mathrm{rods}} = 6$ calculated using FDTD method as implemented in MEEP \cite{MEEP}.}
\label{fig:figure4}
\end{figure*}

When probing reflection through the chiral woodpile, we make the assumption that the sample is effectively infinite in the x and y directions (parallel to the surface) and truncated in the z-direction; we thus examine the projected band structure, since $k_z$ is no longer conserved (but $k_x$ and $k_y$ still are). As such, the projected band structure consists of states with both $k_z = 0$ and $k_z \ne 0$ projected onto the $(k_x,k_y)$ plane. Thus, even though the Weyl point exists in an incomplete band gap in the 3D Brillouin zone of our chiral woodpile PhC, in the projected band structure there are states from other bands which are now degenerate with the Weyl point in frequency and with the same $k_x$ and $k_y$, and one may expect that these overlapping states would obscure the signature of the Weyl point in the reflection spectrum. However, the agreement we observe between the 3D band structure calculations (Fig. \ref{fig:figure1} (b) and dashed lines in Fig. \ref{fig:figure3}) and the reflection spectrum (Fig. \ref{fig:figure3}) suggests that this measurement is relatively insensitive to these overlapping states. To further explore this feature in our results, we calculate the modal overlaps between s- and p-polarized plane waves and the Bloch modes of the PhC. Two parameters are defined using these modal overlaps which measure the polarization of the Bloch modes and the overall in-coupling of an arbitrarily polarized plane wave \cite{CD1, CD2, CD3, CD4}. These parameters indicate that the modes with $k_z \ne 0$ in the projected band structure, that would have otherwise obscured the observation of the Weyl point, either have a polarization mismatch with the incident light and/or have inefficient mode in-coupling. This leads to the observed features in the reflection spectrum wherein the boundaries of highly reflecting regions correspond to the $k_z = 0$ Weyl bands (see supplementary material for a detailed analysis). More generally, this kind of polarization analysis can be useful for any low-contrast PhCs where the projected band structure is insufficient to infer the origins of spectral features from the band structure.

The topological charge associated with Weyl points gives rise to Fermi arc-like surface states which are a direct consequence of the bulk-boundary correspondence. However, at low dielectric contrast, as in the present experiment, the overlapping states and the lack of a bandgap imply that any surface states associated with the Weyl points are leaky resonances that are degenerate with bulk eigenstates of the PhC. Even at high dielectric contrast, the surfaces states associated with the Weyl point at $\Gamma$ are resonances for two reasons: (1) When the structure is truncated, surface states lie above the light line of air and confinement on this side requires a trivial bandgap material and (2) the dispersion of the bands for our structure does not allow the opening of a local bandgap around the Weyl point.  We have found that such resonances (if present) are too broadened by leakage to be observable for the index contrast of the photonic crystal used here.  However, both these issues are resolved when examining a high-contrast version of our PhC, in the vicinity of the charge $-2$ Weyl point at $R$ [$k_x=k_y=k_z=0.5$ ($2\pi/a$)], which we consider for the following numerical analysis.

In a chiral woodpile PhC with $\varepsilon_{\mathrm{rods}} = 12$ and truncated in the z-direction, we examine the surface band structure along a circular loop enclosing the projection of the Weyl point \cite{weylcharge2_phonon, weyl_ss_winding}. This loop has radius $0.15$ $(2\pi/a)$ and is parametrized by an angle $0\le t/2\pi \le 1$. The surface band structure as plotted in Fig. 4(a) reveals the existence of Fermi arc surface states in the non-trivial gap formed along the loop. Due to the bulk-boundary correspondence, the number of these surface states that cross the bulk band gap is equal to the Chern number of the enclosed degeneracy, which in this case is $-2$.  The sign corresponds to the direction of motion of the surface states. The magnetic field intensity of the surface state localized to the top edge is plotted in Fig. 4(b), which shows strong confinement to the PhC-air interface. We repeat these calculations for progressively smaller values of dielectric contrast and find that a local gap fails to open at $\varepsilon_{\mathrm{rods}} \sim 6$ around the loop considered here and at $\varepsilon_{\mathrm{rods}} \sim 5$, a local gap fails to open for a loop of any radius. Examining the field intensity profile from a finite-difference time-domain (FDTD) simulation \cite{MEEP}, shown in Fig. 4(d), reveals that at $\varepsilon_{\mathrm{rods}}=6$, the surface state is now a leaky resonance with a $Q \sim 10^3$. As we can see, the confinement of the surface state relies on the existence of at least a local bandgap, which heavily depends on the dielectric contrast (see animated file in supplementary material).  For the dielectric contrast of our experiment, namely $\varepsilon_{\mathrm{rods}} = 2.31$, the resonances have broadened sufficiently to be effectively unobservable. On the other hand, changing the dielectric contrast does not break the screw symmetry in our structure and as such does not affect the existence of the Weyl points, whether at the $\Gamma$ or $R$-points. Therefore, although the surface states in our low-contrast PhC have turned into resonances that extend into the bulk, the charge $\pm 2$ Weyl points have a continued existence due to their symmetry protection.

To summarize, we have presented the observation of a charge-2 Weyl point in a 3D PhC in the mid-infrared regime. The fact that we have used a relatively low refractive index polymer opens the door to exploring 3D topological phenomena in the IR and into optical frequencies, where very high-index, low-loss materials may not be available or amenable to 3D device fabrication. It will be of interest in the future to explore surface resonances in 3D PhCs with Weyl points; these will have fundamentally different properties compared to traditional surface states because of radiative loss, which will give rise to non-Hermitian effects on the surface.

M. C. R. acknowledges support from the ONR YIP program under grant number N00014-18-1-2595, the Packard Foundation (2017-66821), and the Penn State NSF-MRSEC Center for Nanoscale Science, under award number DMR-1420620. J. N. acknowledges support by Corning Incorporated Office of STEM Graduate Research. We are grateful to Josh Stapleton and Tawanda J. Zimudzi for technical help with the FTIR spectrometer and Kathleen Gehoski for technical help with Nanoscribe. S. V. acknowledges Debadarshini Mishra for helpful discussions on the topic.

S.V. and J.N. contributed equally to this work.
\bibliography{main_charge2_WPs}

\end{document}


\onecolumngrid
\raggedbottom
\begin{center}
    \begin{large}
    \textbf{Supplementary Material for Observation of a Charge-2 Photonic Weyl Point in the Infrared \vskip 12 pt}
    \end{large}
    Sachin Vaidya$^{1}$$^*$, Jiho Noh$^{1}$$^*$, Alexander Cerjan$^{1}$, Mikael C. Rechtsman$^{1}$\\
    $^1$\emph{Department of Physics, The Pennsylvania State University, University Park, PA 16802, USA\\}
    (Dated: February 6, 2020) 
\end{center}

\author{Sachin Vaidya$^{1}$$^*$, Jiho Noh$^{1}$$^*$, Alexander Cerjan$^{1}$, Mikael C. Rechtsman$^{1}$}
\affiliation{
 $^1$Department of Physics, The Pennsylvania State University, University Park, PA 16802, USA
}

\def\thefootnote{*}\footnotetext{These authors contributed equally to this work}

\section{Band Structure}
\begin{figure}[H]
\centering
\includegraphics[scale = 0.175]{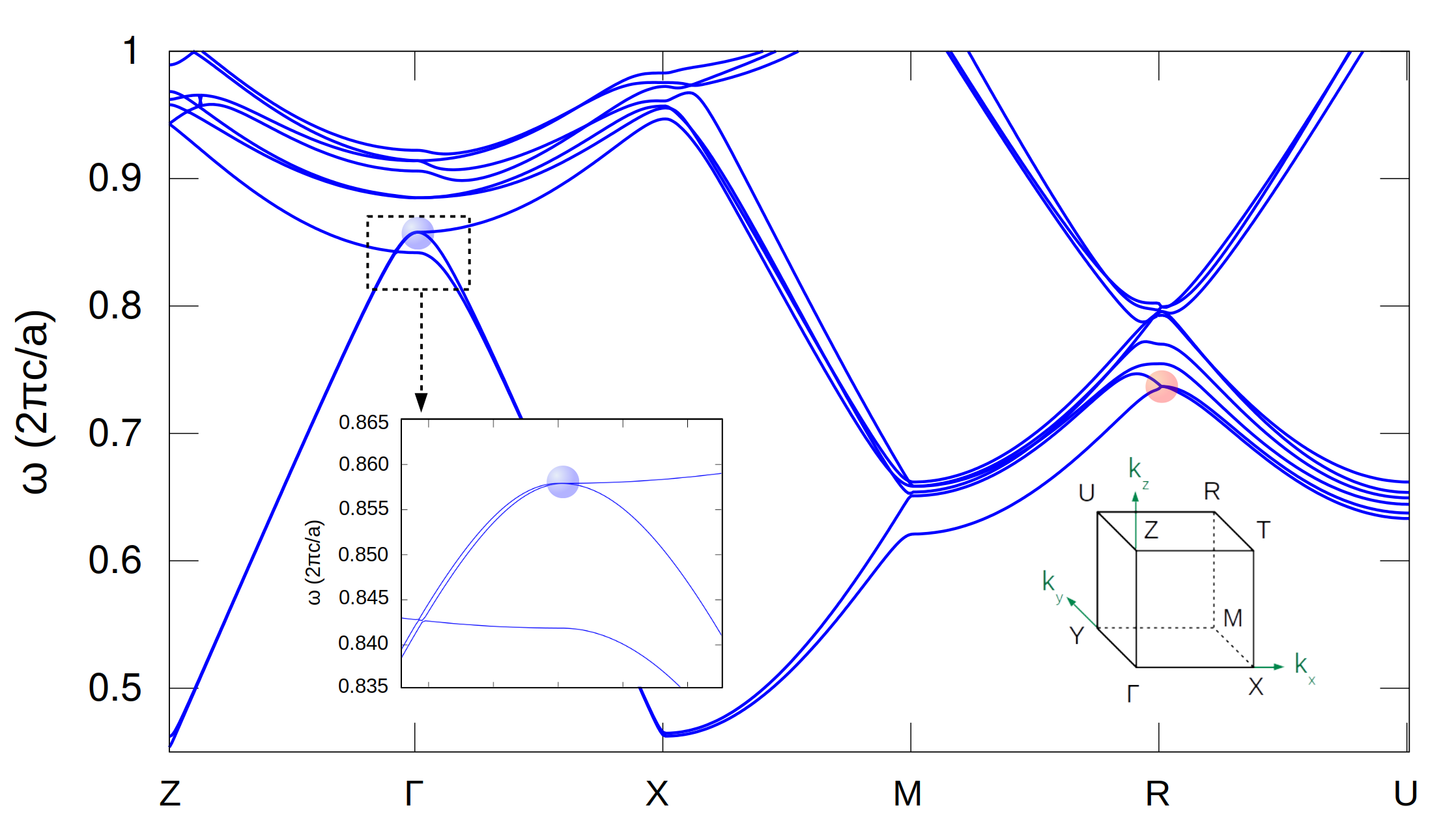}
\caption{\textbf{3D band structure of the chiral woodpile PhC.} The PhC has the following parameters: dielectric constant of rods ($\varepsilon_{\mathrm{rods}})$ $=2.31$, rod width $(w) = 0.175a$ and rod height $(h) = 0.25a$, where $a$ is the lattice constant in all three directions. The inset contains the bands in the vicinity of $\Gamma$. The charge $\pm2$ Weyl points are marked with a blue and red circle respectively.}
\label{fig:figure1s}
\end{figure}

The full 3D band structure for the chiral woodpile structure discussed in the main text is plotted in Fig. \ref{fig:figure1s} The two bands that make up the Weyl point are well-separated in frequency along the $Y$$-\Gamma-$$X$ directions but are very close in frequency along $\Gamma-Z$. We perform convergence tests in MPB that confirm that the degeneracy only occurs at the $\Gamma$ point. This structure also has a charge $-2$ Weyl point at R between bands 3 and 4 which lies below the light line of air.

\section{Calculation of Berry Phase}
\begin{figure}[H]
\centering
\includegraphics[scale = 0.225]{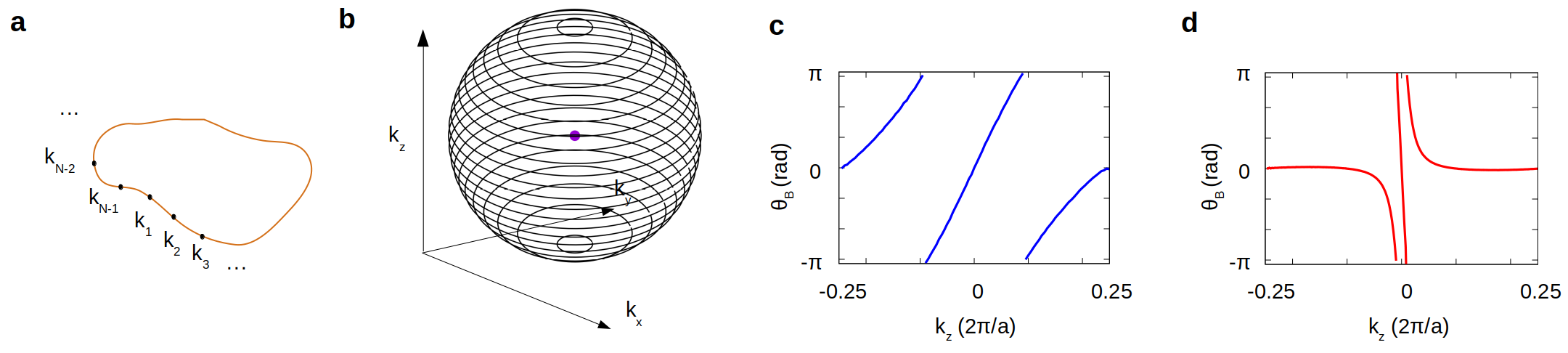}
\caption{\textbf{Calculation of the Berry phase around a Weyl point.} (a) A closed contour in k-space. (b) The Berry phase is calculated on circular contours that enclose the Weyl point. Each contour lies on a constant $k_z$ plane. (c), (d) Berry phase vs. $k_z$ plots for charge $+2$ and charge $-2$ Weyl points at $\Gamma$ and R respectively. }
\label{fig:figure2s}
\end{figure}

As discussed in the main text, we have numerically calculated the topological charge (Chern number)  associated with the Weyl points at $\Gamma$ and $R$ of the chiral woodpile PhC. Here we describe that algorithm, following Ref. \cite{resta}. Consider a closed contour $C$ in momentum ($\mathbf{k}$) space discretized into $N$ points $(\mathbf{k_1}, ..., \mathbf{k_{N-1}})$ such that $\mathbf{k_N} = \mathbf{k_1}$. The total phase acquired by a state undergoing adiabatic transport along this loop is numerically calculated by
\begin{equation}
    \theta_B(C) = - \textrm{Im} \ln\left(\prod_{i = 1}^{N}\frac{\braket{\psi(\mathbf{k_{i}})|\psi(\mathbf{k_{i+1})}}}{\left|\braket{\psi(\mathbf{k_{i}})|\psi(\mathbf{k_{i+1}})}\right|} \right)
    \tag{S1}
\end{equation}
$\theta_B(C)$ is a manifestly gauge-invariant quantity and is called the Berry phase.
In our case, the states in a 3D PhC are magnetic field eigenmodes of the Maxwell operator \cite{photoniccrystalsbook} that satisfy 
\begin{equation}
    \nabla \times \left( \frac{1}{\varepsilon(\mathbf{r})} \nabla \times \mathbf{H}(\mathbf{r}) \right) = \left(\frac{\omega}{c}\right)^2 \mathbf{H}(\mathbf{r}) \tag{S2} 
    \end{equation}
    and the inner product between two such eigenmodes is defined as \begin{equation}
    \braket{\psi_1|\psi_2} = \int \mathbf{H}^*_{1}(\mathbf{r})\cdot \mathbf{H}_{2}(\mathbf{r}) d^3\mathbf{r} \tag{S3} \end{equation}
This phase is calculated on circular contours that lie on constant $k_z$ planes as shown in Fig. \ref{fig:figure2s} (b). The topological charge of the Weyl point is the number of times the Berry phase winds as a function of $k_z$.
To show that the Weyl points are indeed symmetry protected and continue to exist even at low contrast, we start with a high contrast version of our PhC and plot the band structure and Berry phase as the contrast is smoothly lowered. An animated version of these plots can be found in the supplementary material.

\section{Coupling Coefficients Between Incident Plane Waves and PhC bloch modes}
To understand the relationship between the observed spectral features and the bulk bands, we calculate coefficients that measure the polarization of Bloch modes. For this, we consider plane waves with s- and p- polarizations incident along the $\Gamma - X$ direction on the PhC-air interface and define the polarization coupling coefficients $C_s$ and $C_p$ and mode coupling index $\kappa$ \cite{CD1, CD2, CD3, CD4}
\begin{equation}
    \eta_{p, \mathbf{k},n} = \frac{\left|\iiint \mathbf{\hat{y}}\cdot \mathbf{H}_{\mathbf{k},n}(x,y,z)dxdydz \right|^2}{V \iiint \left| \mathbf{H}_{\mathbf{k},n}(x,y,z)\right|^2dxdydz} 
\end{equation}

\begin{equation}
    \eta_{s, \mathbf{k},n} = \frac{\left|\iiint (\cos\theta \mathbf{\hat{x}} - \sin\theta\mathbf{\hat{z})}\cdot \mathbf{H}_{\mathbf{k},n}(x,y,z)dxdydz \right|^2}{V \iiint | \mathbf{H}_{\mathbf{k},n}(x,y,z)|^2dxdydz} 
\end{equation}

\begin{equation}
    \kappa = \eta_{s,\mathbf{k},n} + \eta_{p,\mathbf{k},n}; \quad C_{s/p} = \frac{\eta_{s/p,\mathbf{k},n}}{\kappa}
\end{equation}

where $\mathbf{H}_{\mathbf{k},n}(x,y,z)$ is the magnetic field eigenmode of the PhC with momentum $\mathbf{k}$ and band index $n$, $\theta$ is the angle of incidence along $\Gamma - X$ and all integrals are calculated over a unit cell with volume $V$. $\kappa$ goes from 0 to 1 and measures the strength of coupling to a plane wave of any arbitrary polarization while $C_s$ and $C_p$ measure the overlap between s- and p-polarized plane waves and the Bloch modes. Large reflection for a certain range of angles and frequencies in the absence of band gaps can then be thought of as either a polarization mismatch between the incident wave and the Bloch mode, indicated by small $C_{s/p}$, and/or inefficient mode in-coupling, indicated by small $\kappa$. 

\begin{figure}[H]
\centering
\includegraphics[scale = 0.45]{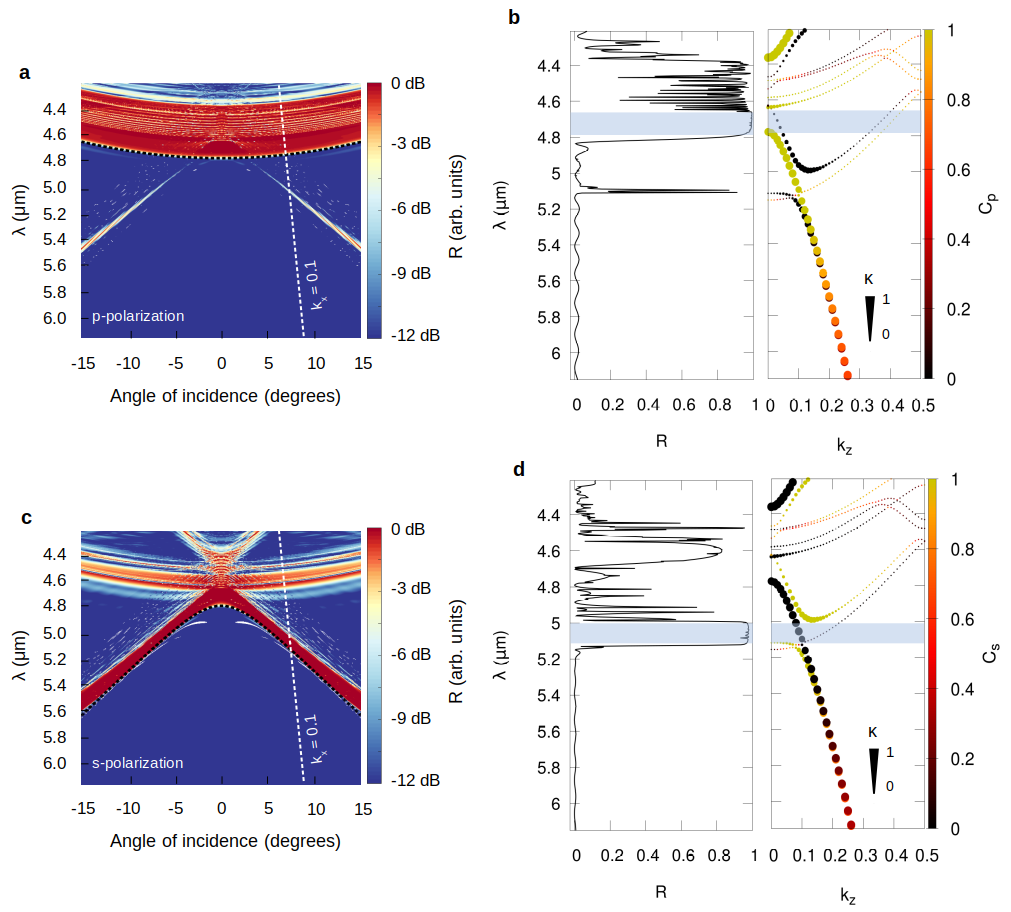}
\caption{\textbf{Polarization analysis of Bloch modes}. (a), (c) Angle-resolved RCWA simulation of reflection spectra for p and s polarizations respectively as shown in Fig. 3 of the main text. (b), (d) Reflection spectrum along the cut shown by the dashed line in (a) and (c) and the band structure showing bands 3 to 12 for $\mathbf{k} = (0.1,0,0)$ to $(0.1,0,0.5)$ ($2\pi /a$). The color of the circular dots indicates the value of $C_{s/p}$ for the corresponding Bloch modes and their size is proportional to the value of $\kappa$.}
\label{fig:figure3s}
\end{figure}

In Fig. \ref{fig:figure3s}, we analyze a slice of the reflection spectrum that corresponds to Bloch modes with momenta $\mathbf{k} = (0.1,0,k_z)$ ($2\pi /a$). As previously stated, the band structure weighted by the coupling coefficients shows that there are wavelength ranges where the states are nearly completely s- or p- polarized and/or have small mode coupling index. We particularly point out the wavelength ranges $\sim 4.65-4.8$ $\mu m$ in Fig. \ref{fig:figure3s} (b) and $\sim 5.0-5.15$ $\mu m$ in Fig. \ref{fig:figure3s} (d) (highlighted in blue) which coincide with the sharp increases in reflection. Moreover, the boundary separating the reflecting and transmissive region coincides with band 4 at $k_z = 0$ in s-polarization and band 5 at $k_z = 0$ in p-polarization plots, which are the Weyl bulk bands of interest. Thus the signatures of both bands and the Weyl point are present in simulations and measurements done with $45^{\circ}$ polarization.

\bibliography{supplement_charge2_WPs}